# Electronic Phonons in a Moiré Electron Crystal

Yan Zhao[1,2†], Yuhang Hou[3†], Xiangbin Cai[1,2†], Shihao Ru[1,2,4], Shunshun Yang[1,5], Yan Zhang[6], Xuran Dai[1,2], Qiuyu Shang[1,2], Abdullah Rasmita[1,2], Haiyang Pan[1,2], Kenji Watanabe[7], Takashi Taniguchi[8], Hongbin Cai[9], Hongyi Yu[3,10*], Weibo Gao[1,2,4,11*]

[1]*Division of Physics and Applied Physics, School of Physical and Mathematical Sciences, Nanyang Technological University, Singapore 637371, Singapore*

[2]*School of Electrical & Electronic Engineering, Nanyang Technological University, Singapore 639798, Singapore*

[3]*Guangdong Provincial Key Laboratory of Quantum Metrology and Sensing & School of Physics and Astronomy, Sun Yat-Sen University (Zhuhai Campus), Zhuhai, China*

[4]*Centre for Quantum Technologies, Nanyang Technological University, Singapore, Singapore*

[5]*School of Physics, Nanjing University of Aeronautics and Astronautics, Nanjing, China*

[6]*Center for artificial muscles, École Polytechnique Fédérale de Lausanne (EPFL), 2002 Neuchâtel, Switzerland*

[7]*Research Center for Electronic and Optical Materials, National Institute for Materials Science, 1-1 Namiki, Tsukuba 305-0044, Japan*

[8]*Research Center for Materials Nanoarchitectonics, National Institute for Materials Science, 1-1 Namiki, Tsukuba 305-0044, Japan*

[9]*Hefei National Research Center for Physical Sciences at the Microscale, University of Science and Technology of China, Hefei, China.*

[10]*State Key Laboratory of Optoelectronic Materials and Technologies, Sun Yat-Sen University (Guangzhou Campus), Guangzhou, China*

[11]*Quantum Science and Engineering Centre (QSec), Nanyang Technological University, Singapore, Singapore*

[†]*These authors contributed equally: Yan Zhao, Yuhang Hou, Xiangbin Cai*

[*]*Correspondence to H.Y. (yuhy33@mail.sysu.edu.cn) and W.G. (wbgao@ntu.edu.sg).*

**Abstract**

**Collective quantum phenomena, such as the excitation of composite fermions[1], spin waves[2], and exciton condensation[3,4], can emerge in strongly correlated systems like the fractional quantum Hall states[5], spin liquids[6], or excitonic insulators[7]. Two-dimensional (2D) moiré superlattices have emerged as a powerful platform for exploring such correlated phases and their associated collective excitations[8,9]. Specifically, electron crystals stabilized by long-range Coulomb interactions may host collective vibrational excitations emerging from**

**electron correlations[10], termed electronic phonons, which are fundamentally distinct from atomic lattice phonons. Despite theoretical prediction of their existence in moiré electron crystals[11], direct experimental evidence has remained elusive. Here we report the observation of electronic phonons in the Mott insulating and stripe phases of a $WS_2/WSe_2$ moiré superlattice, achieved through light scattering measurements. The phonon energies, temperature and filling factor dependencies, along with theoretical modeling, corroborate their origin as collective vibrations of a correlated electron crystal. Polarization-resolved measurements further indicate rotational symmetry breaking in the Mott state. Notably, these electronic phonons exhibit strong tunability in energy, intensity, and polarization under external electric or magnetic fields, highlighting rich and controllable lattice dynamics of the electron crystal. These findings provide direct spectroscopic evidence for the electronic crystalline nature of correlated phases, opening avenues for probing and manipulating collective excitations in correlated electron systems.**

**Main Text**

Collective excitation is a fundamental concept in solid-state physics and often arises as the collective oscillations of order parameters within quantum materials. Various types of elementary excitations have been observed, including magnons[12-14] from spin waves, amplitudons and phasons[15-17] from charge density waves, orbitons[18] linked to orbital ordering and gravitons[5] associated with quantum metric fluctuations. Candidate materials that support these excitations include magnets, superconductors[19,20], and topological systems such as the fractional quantum Hall insulators[21-23], where the emergence of collective excitations and electronic phase transitions is often intimately tied to symmetry breaking. Understanding and controlling collective quantum excitations are essential for unraveling the complex interactions underlying intertwined electronic orders and strongly correlated phases.

Two-dimensional (2D) moiré superlattices have emerged as a versatile platform for generating a rich variety of correlated quantum phases, including Mott insulators[24,25], generalized Wigner crystals[26,27], and topologically nontrivial quantum anomalous Hall states[28-30]. Notably, the enhanced Coulomb interactions and moiré potential confinement can stabilize electron crystals at both integer and fractional fillings[27,31,32], in contrast to conventional Wigner crystals that require high magnetic field to suppress electronic kinetic energy. A hallmark of these long-range ordered electronic solids is the emergence of collective vibrational excitations[10], where electrons no longer behave as isolated particles but oscillate collectively as a coherent many-body entity—akin to phonons in atomic lattices, which can be termed electronic phonons. Unlike conventional phonons from atomic displacements, they originate from electronic correlations and are intrinsically more tunable by external fields due to the smaller effective mass and charged nature of the electron constituents. In high-field GaAs Wigner crystals, signatures of collective vibrations were indirectly inferred from tunnelling measurements[33]. However, the requirement of strong magnetic fields, ultralow temperatures, and low carrier densities presents challenges for the direct spectroscopy observation. In contrast, moiré superlattices provide a highly tunable, magnetic-field-free platform to directly probe and control electronic phonons. Yet, despite theoretical anticipation[11], direct spectroscopic detection and manipulation of such excitations in moiré electron crystals has remained elusive.

Here we report the direct observation of electronic phonons in the correlated phases of a $WS_2/WSe_2$ moiré superlattice thanks to the enhanced exciton-phonon interaction, using polarization-resolved Raman spectroscopy. Distinct Raman modes emerge at both integer and fractional fillings, with the strongest features appearing at the Mott insulating state ($v = 1$). Combined experimental and theoretical evidence suggests these are phonons of the electron crystal, supported by their temperature dependence, filling factor evolution, symmetry properties, and modeling. Remarkably, these phonons exhibit significant in-plane anisotropy that is tunable via an out-of-plane magnetic field, indicating a breaking of rotational symmetry in the electron crystal. Furthermore, the energies and polarizations of the electronic phonons exhibit tunability under an out-of-plane electric field, and their polarizations are also sensitive to an applied magnetic field. These findings provide spectroscopic evidence of lattice dynamics in a correlated electron crystal, establishing moiré materials as a platform for exploring and controlling collective quantum excitations in strongly interacting systems.

**Observation of electronic phonons in correlated phases**

We fabricated $WS_2/WSe_2$ moiré heterostructure (see Methods for details) encapsulated by hexagonal boron nitride (hBN) and contacted with graphite gates, enabling dual-gate control over the carrier density ($n$) and vertical electric field ($E$) (Fig. 1a). To identify correlated electronic phases, we monitored the interlayer exciton (IX) photoluminescence (PL), of which both the energy and intensity are highly sensitive to the formation of correlated phases[34]. As shown in Fig. 1b, the doping-dependent IX PL spectra reveal distinct filling factors ($v$) through enhanced intensity and blueshift (see Extended Data Fig. 1). In particular, correlated insulating and charge-ordered phases are identified, including the Mott insulating states at $v = -1$ and $1$ ($n \sim 2.0 \times 10^{12}$ cm$^{-2}$) and generalized Wigner crystal states at $v = -1/2, 1/2$, and $3/2$ (see Supplementary Section 1 for details). Unless otherwise specified, all measurements were performed at 5 K using an excitation laser energy of $E_L$=1.96 eV.

Figure 1c shows the Raman spectra as a function of carrier density. The prominent peak near 420 cm$^{-1}$ corresponds to the $A_{1g}$ phonon mode of the WS$_2$ atomic lattice, which diminishes with increasing electron doping (the Raman signals of WS$_2$ are much stronger than those of WSe$_2$ due to resonant enhancement under 1.96 eV excitation for WS$_2$ but not WSe$_2$[35]). Notably, two additional peaks—labeled M and N for convenient reference—emerge within specific carrier density ranges, with their peak energies varying at different fillings. This is clearly illustrated in the extracted Raman spectra at selected filling factors (Fig. 1d). Peak M appears at 405 cm$^{-1}$ ($v = 1/2$), 406 cm$^{-1}$ ($v = 1$) and 415 cm$^{-1}$ ($v = 3/2$), while peak N is located at 503 cm$^{-1}$ ($v = 1/2$) and 517 cm$^{-1}$ ($v = 1$), but is not clearly resolved at $v = 3/2$. Figure 1e summarizes the integrated intensities of M and N as a function of $n$. It's clear that both modes are confined to narrow density ranges centered around correlated fillings, with the strongest response at the Mott state ($v = 1$), and weaker intensities at fractional fillings ($v = 1/2$ or $3/2$). Figure 1f compares the intensity maps of IX PL and peak M as functions of $n$ and $E$. The distinct vertical features in the PL map correspond to the enhanced emission at specific filling factors, while the Raman map highlights the prominent intensity at $v = 1$ across a wide electric-field range and indicates its robust correlation with the charge-ordered state (the electric-field dependence is discussed further below).

To investigate whether the emergence of the new Raman peaks is tied to electron crystallization, we measured the temperature dependence of peak M at fixed filling factor $v = 1$, as shown in Fig. 2a and b. Peak M persists up to ~190 K. The critical temperature ($T_c$) for peak N is similar (Supplementary Fig. 1). At fractional fillings $v = 1/2$ and $3/2$, peak M vanishes at lower temperatures, with $T_c$ values around 50 K and 90 K, respectively (Extended Data Fig. 2). These thermal melting behaviors are consistent with those of the correlated phases evaluated through the IX PL (Extended Data Fig. 3). The relatively high $T_c$ values reported here, compared to those assigned by electrical transport or optical methods in previous studies[32,36], may reflect the differences in sensitivity between experimental techniques or variations in sample quality and device configurations.

To further probe the nature of these new peaks, we studied their excitation energy dependence (Fig. 2c; see also Supplementary Fig. 2). First, the spectral position of peak M remains constant across different excitation energies, ruling out PL origins and supporting its identification as a collective mode (the same holds for peak N; see Supplementary Fig. 3). Second, the intensity of peak M exhibits pronounced resonant enhancement effect with the maximum appearing around the excitation energy of 1.975 eV (Fig. 2d), which is close to the 1$s$ intralayer exciton of WS$_2$ (see absorption spectrum in Supplementary Fig. 4). This behavior indicates a strong coupling between the collective mode and the WS$_2$ exciton. It is noteworthy that these new peaks are only observed in the electron-doped regime, where the electron crystal forms in the WS$_2$ layer and interacts strongly with the intralayer exciton. Although interlayer hybridization may occur, the spectroscopic signatures indicate that the collective modes observed here are primarily associated with the electrons in WS$_2$. Taken together, the temperature, filling factor, and excitation energy dependencies strongly suggest that the observed new peaks arise from collective excitations of the electron crystal. Their identification as electronic phonons is further supported by theoretical modeling, as detailed in Fig. 4. In the following text, we focus on the detailed characteristics of these modes at $v = 1$, where the signal is most pronounced.

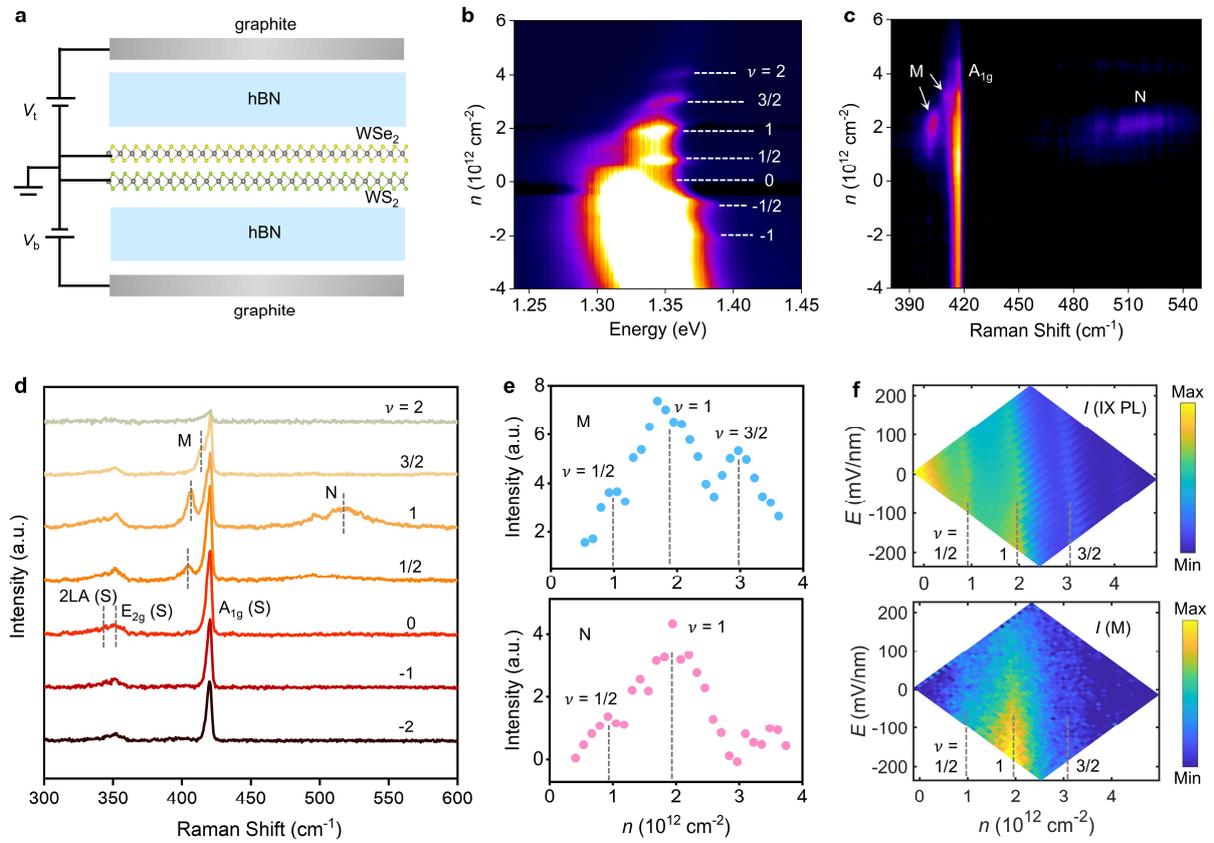

**Fig. 1 | Observation of electronic phonons in correlated phases. a**, Schematic of the WS$_2$/WSe$_2$ moiré heterostructure device, with dual graphite gates and hBN dielectric layers. **b**, Interlayer exciton (IX) photoluminescence (PL) spectra as a function of carrier density ($n$). Filling factors are illustrated by dashed lines. **c**, Raman spectra as a function of $n$, with $E$ fixed at zero. Two new Raman peaks are labeled as M and N. **d**, Raman spectra at selected filling factors. Atomic lattice phonon modes of WS$_2$ are marked by (S). **e**, Integrated Raman intensities of M (top panel) and N (bottom panel) versus $n$. **f**, False-color intensity maps of IX PL (top panel) and peak M (bottom panel) as functions of $n$ and $E$, showing a strong response of M around the Mott insulating state ($v = 1$).

**Symmetry analysis by polarization-resolved measurements**

To elucidate the symmetry properties of the observed electronic phonons, we investigated their polarization selection rules using polarization-resolved Raman spectroscopy in backscattering geometry. We first fixed the incident light polarization and analyzed the scattered light under both linear and circular configurations. As shown in Extended Data Fig. 4, both peaks M and N are co-polarized with the incident light, analogous to the selection rule of $A_{1g}$ mode in $WS_2$. However, a more detailed polarization analysis revealed key differences. By continuously rotating the analyzer while keeping the incident linear polarization fixed (Fig. 2e), we observed a distinct phase shift ($\Delta\theta \sim 17°$) between the polarization states of peak M and $A_{1g}$ mode (see Supplementary Fig. 5 for phase differences at additional positions). This phase shift, also seen for peak N (Supplementary Fig. 6), is evident in the polar plots shown in Fig. 2g. Similar polarization deviation between M and $A_{1g}$ also occurs for the circularly polarized configuration (Supplementary Fig. 7). The $A_{1g}$ mode arises from the out-of-plane vibrations of sulfur atoms[37] and strictly preserves the polarization of incident light, as expected for a rotationally symmetric vibrational mode. The polarization deviation of peak M indicates that it generates a polarization component orthogonal to that of the incident light, which could be induced by the competition between an intrinsic polarization direction from the in-plane anisotropy of the electron crystal and the polarization of the incident light.

To verify whether there is an in-plane anisotropy, we further rotated the linear polarization direction of the incident light and analyzed the scattered intensity (Fig. 2f). As shown in Fig. 2h, while the $A_{1g}$ mode remains isotropic, the intensities of peaks M and N exhibit a pronounced anisotropy (similar anisotropy also appears in device D2, see Extended Data Fig. 5). This confirms that the symmetry of the electronic phonons differs fundamentally from the atomic phonons, with the former lacking the threefold rotational symmetry ($C_3$) of the underlying lattice. The observed in-plane anisotropy suggests that the electron crystal exhibits a preferred direction, implying either extrinsic symmetry breaking—such as from strain—or spontaneous electronic nematicity. At fractional fillings like $v = 1/2$ and $3/2$, stripe phases are expected and naturally break rotational symmetry[31]. However, the electron crystal at $v = 1$ is generally predicted to retain $C_3$ symmetry in the absence of extrinsic strain or disorder effects[26]. Although some degree of strain is unavoidable during device fabrication and some kinds of symmetry breaking may be introduced, they remain below the detection threshold of polarization-dependent PL spectra or linear dichroism measurements (Extended Data Fig. 6), which are sensitive to the symmetry breaking in the atomic lattice[38,39].

Two scenarios may account for the observed anisotropy: (1) the in-plane symmetry breaking is introduced by external perturbation such as strain, although it is too small to be sensed by the atomic lattice, the moiré pattern can magnify its effect through the distortion of the electron crystal[40]; (2) the symmetry breaking arises spontaneously from electron-electron interactions, leading to an anisotropic electronic order such as a nematic phase. Notably, similar electronic nematicity has been observed at half-filling Mott insulator state in twisted bilayer graphene using scanning tunnelling spectroscopy[41], although its origin remains debated. In our experiment, although the microscopic origin of the symmetry breaking cannot be unambiguously identified at present (see further discussion in Supplementary Information Section 2), the simultaneous emergence of the new mode and its polarization anisotropy in multiple devices suggests that the reduced symmetry is intimately linked to this collective electronic excitation, which warrants further experimental and theoretical investigation.

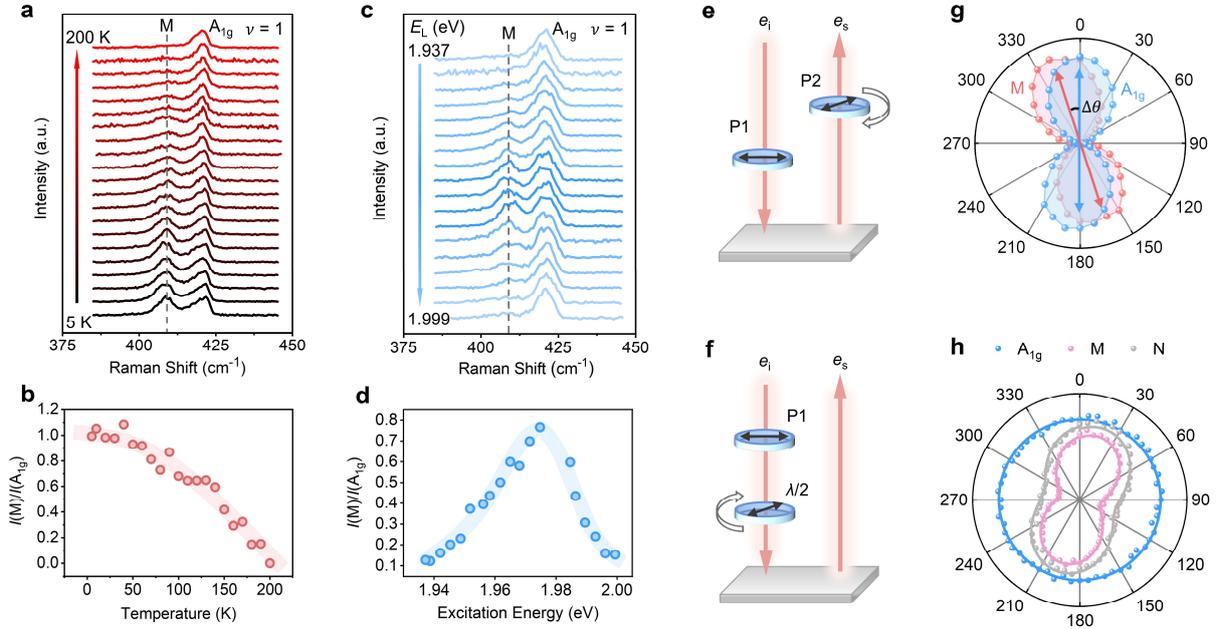

**Fig.2 | Temperature, excitation energy, and polarization dependence of electronic phonons in the electron crystal. a**, Evolution of Raman spectra at $\nu = 1$ with temperature. Each spectrum is normalized to the intensity of the $A_{1g}$ mode. **b**, Intensity ratio of peak M to the $A_{1g}$ mode as a function of temperature. Peak M disappears above 190 K. Pink shade is guide to the eye. **c**, Raman spectra at $\nu = 1$ with the variation of excitation laser energy, showing the evolution of peak M. **d**, Resonant profile of peak M, plotted as the intensity ratio of M to $A_{1g}$ versus excitation energy. A pronounced resonance is observed near 1.975 eV, close to the 1$s$ exciton energy of $WS_2$. Blue shade is guide to the eye. **e,f**, Schematic of linear polarization-resolved measurement configurations: (**e**) the incident polarization P1 is fixed while the analyzer P2 is rotated; (**f**) the incident polarization P1 is continuously rotated using a half-wave plate ($\lambda/2$). $e_i$ and $e_s$ represent

the incident and scattered lights, respectively. **g**, Polarization states of peaks M and $A_{1g}$ (arrows mark the polarization directions) measured by the optical configuration in (**e**), showing a distinct phase difference $\Delta\theta$. **h**, Intensity polar plots as a function of incident polarization angle for peaks M, N, and $A_{1g}$, obtained from the configuration in (**f**). Peaks M and N exhibit strong in-plane anisotropy, in contrast to the isotropic $A_{1g}$ phonon.

**Electric and magnetic field modulation**

Next, we focus on the dynamics of electronic phonons under external fields. We first examine the effect of an applied out-of-plane electric field. The dual-gate device configuration allows independent tuning of the vertical electric field while maintaining a constant carrier density. Fig. 3a presents the Raman spectra at $v = 1$ under varying electric field. A clear modulation of Raman intensities is observed, particularly in the relative strengths of the M and $A_{1g}$ modes. As shown in Fig. 3b, both M and $A_{1g}$ intensities decrease as the electric field is tuned from negative ($V_t < V_b$, $E$ pointing from $WS_2$ to $WSe_2$) to positive ($V_t > V_b$, $E$ pointing from $WSe_2$ to $WS_2$) values. However, M exhibits a much steeper decline than $A_{1g}$, resulting in a monotonic decrease in the intensity ratio $I(M)/I(A_{1g})$. We attribute the electric field dependence of $I(M)/I(A_{1g})$ to the following reasons: (1) The modulated exciton-phonon interaction[42,43], which is a crucial process in Raman scattering[44] and known to be tunable by the electric field[45]. The electric field can affect the exciton through changing the electron and hole wave function distributions in the out-of-plane direction, which can change both the exciton oscillator strength and the exciton-phonon interaction. Given the pronounced resonant enhancement of peak M under excitonic excitation (Fig. 2c, d), exciton effects are expected to play a more profound role in the electric field response of M compared to $A_{1g}$. (2) Unlike phonons in atomic crystals, the electronic phonon (schematically illustrated in the inset of Fig. 3b) is itself susceptible to the electric field, introducing an additional layer of complexity to the field response. The field modulation on the electronic phonons is directly reflected in their peak energies. As shown in Fig. 3c, the peak position of $A_{1g}$ remains constant while both M and N show clear energy shifts—especially pronounced for N (see more discussions in Supplementary Section 3). The intensity and energy variations of electronic phonons can be related to the field-modulated electron-electron interaction and moiré confinement strength, as will be discussed in the next section.

The polarization behavior of peak M is also found to vary with the electric field. As displayed in Fig. 3d (measured using the polarization-resolved configuration in Fig. 2e), the polarization angle of M mode exhibits a measurable phase shift with changing electric field, whereas the $A_{1g}$ mode remains unaffected. The extracted phase angles ($\theta$) of these two modes and their phase difference ($\Delta\theta$) are plotted in Fig. 3e, revealing a continuous modulation of the M-mode polarization with a total phase change of ~8° across the applied field range. Intensity polar plots (similar to Fig. 2h) under different electric fields show that the in-plane anisotropy declines with increasing electric field, with its direction remaining the same (see Extended Data Fig. 7). This naturally reduces $\Delta\theta$ which should be 0 when the anisotropy vanishes. Other mechanisms that affect the interplay between the in-plane anisotropy and the incident light polarization could also modify $\Delta\theta$, e.g., a modulation to the electron-electron interaction strength (see Supplementary Section 4). All behaviors above give a direct manifestation of the electric field tunability to the electronic phonons.

We next examine how an out-of-plane magnetic field influences the electronic phonons. Notably, the energies of peaks M and N remain unchanged under varying magnetic field (see Extended Data Fig. 8). However, the polarization states of them are remarkably modulated (see Extended Data Fig. 9), manifested as magneto-optical Raman effect, a phenomenon akin to the Kerr or Faraday effects, where the magnetic field interacts with light through the medium to rotate its polarization state[46]. Additionally, the in-plane anisotropy of peak M is found to be tunable by the applied magnetic field. Fig. 3f displays the intensity polar plots of peak M with rotating the incident linear polarization (Fig. 2f), which exhibits a clear shift in the angle of maximum intensity with magnetic field. For an electron crystal lacking $C_3$ but having a vertical mirror plane, the shift of the maximum intensity angle can be qualitatively understood from the following symmetry consideration. When without the magnetic field, the intensity major axis of the electronic phonon mode must be along or perpendicular to the mirror plane. The applied out-of-plane magnetic field breaks the vertical mirror symmetry, resulting in the rotation of the intensity major axis. Note that a vertical mirror reflection transforms the magnetic field to its opposite value, which then leads to the opposite shifts for positive and negative field values. A microscopic analysis for this magnetic field dependence will be given in the next section. This behavior reveals that the in-plane anisotropic response of the electron crystal can be modulated via a magnetic field, offering a potential route to control the optical and electronic properties of the correlated charge order.

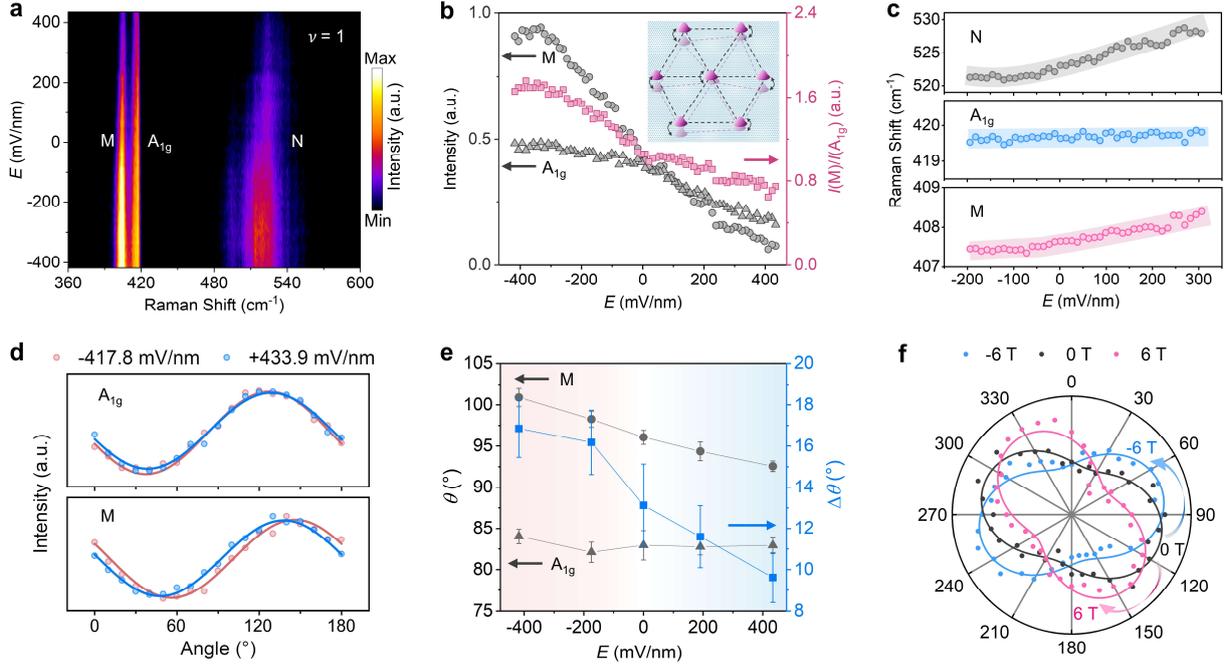

**Fig. 3 | Modulation of electronic phonons by electric and magnetic fields. a**, Raman spectra at $\nu = 1$ with the variation of electric field. **b**, Intensities of M and $A_{1g}$ modes, and their intensity ratio $I(M)/I(A_{1g})$, as a function of electric field. The inset schematically illustrates the electronic phonon in a moiré electron crystal. **c**, Peak positions of M, N and $A_{1g}$ as functions of electric field, showing field-induced energy shifts for M and N. **d**, Polarized Raman intensities of M and $A_{1g}$ measured at electric fields of −417.8 mV/nm and 433.9 mV/nm using the optical configuration in Fig. 2e, demonstrating electric-field-induced modulation of polarization states. **e**, Extracted polarization phase angles ($\theta$) of M and $A_{1g}$, and their difference ($\Delta\theta$), plotted as a function of electric field. **f**, Polar intensity plots of peak M under different magnetic fields, obtained by rotating the linear polarization of the incident light. The orientation of maximum intensity shifts with field, showing opposite rotation directions for positive and negative magnetic fields relative to zero field.

**Theoretical model**

To further confirm the electronic-phonon origin of the newly observed peaks and to understand their anisotropic behavior in the electron crystal, we employ a simplified model based on the interaction between excitons and electronic phonons. We consider a triangular-type electron crystal shown in Fig. 4a. To account for the experimentally observed in-plane anisotropy, the electron crystal is assumed to be expanded along $y$ direction by a small fraction $\eta$. The long-range Coulomb interaction between different electron sites leads to the emergence of electronic phonons, characterized by the frequency $\omega_{l,\mathbf{k}}$ and the phonon polarization[11]. Here $\mathbf{k}$ is the wave vector and $l$ = 1, 2 denotes the two phonon branches. Fig. 4b shows the calculated $\omega_{l,\mathbf{k}}$ and the corresponding phonon polarization for an electron crystal under $\eta = 2\%$, $\lambda_x = 8$ nm and an electron effective mass $m_e = 0.75 m_0$ ($m_0$ corresponds to the free electron mass). A detailed derivation process for the electronic phonon Hamiltonian $\widehat{H}_{\text{ph}}$ and other simulation parameters are provided in Supplementary Section 3. The calculated electronic phonon dispersions of $v = 1/2$ and $v = 3/2$ are shown in Supplementary Section 5.

The electron crystal can be viewed as inversion-symmetric at a moiré length scale ~ 8 nm, and the electronic phonon at $\mathbf{k} = 0$ has an odd parity thus should be Raman-inactive. Therefore, we attribute the Raman peaks M and N to two-phonon emissions mediated by the exciton, with the whole process schematically illustrated in Fig. 4c. Here the electron-hole exchange interaction splits the exciton into the transverse and longitudinal branches at a finite center-of-mass momentum. Then we use an interaction Hamiltonian between the exciton and electronic phonons (see details in Methods) to simulate the following scenarios performed in experiments: (1) under a circularly polarized incident light, the co-circularly (cross-circularly) polarized Raman intensity $I_+$ ($I_-$) is measured as a function of the Raman shift $\omega = \omega_{l,-\mathbf{k}} + \omega_{l',\mathbf{k}}$; (2) under a linearly polarized incident light, the co-linearly (cross-linearly) polarized Raman intensity $I_\parallel$ ($I_\perp$) is measured, and the polar plot of $I_\parallel + I_\perp$ as a function of the incident polarization direction is analyzed to assess the in-plane anisotropy; (3) an out-of-plane magnetic field is applied to explore its effect on the rotation of the maximum intensity orientation in the polar plot of $I_\parallel + I_\perp$.

Fig. 4d and 4e show the simulation results of the polarization-resolved Raman spectra (see Supplementary Section 6 for details). We note that the quantitative line shape of the Raman spectrum varies with system parameters including the adopted Coulomb interaction form, the moiré confinement strength, the exciton dispersion relation, etc. Despite being an oversimplified model, our simulation can well reproduce most of the qualitative features observed in experiments:

1) There always exist a lower-energy strong peak and a higher-energy weak peak for all sets of the considered parameter values, which can be tuned to near 407 cm$^{-1}$ and 523 cm$^{-1}$ thus are attributed to peaks M and N, respectively. They arise from the enhanced phonon density of states associated with van Hove singularities in the phonon dispersion (M point of the Brillouin zone). The experimentally observed difference in linewidth between peaks M and N is discussed separately in Supplementary Section 7.
2) The two peaks exhibit strongly co-polarized Raman emissions under both circularly and linearly polarized excitations (see Fig. 4d). Notably, when the linear polarization of the incident light is not parallel/perpendicular to the anisotropy axis, a finite phase difference $\Delta\theta$ between polarization states of the Raman and incident light can be obtained. Furthermore, our simulation indicates that $\Delta\theta$ can be modulated by varying the electron-electron interaction strength, even if the in-plane anisotropy is not changed (see Supplementary Fig. 9).

3) The intensities and energies of the two peaks can be significantly changed by varying the electron-electron interaction and moiré confinement strengths. The inset in the left panel of Fig. 4d shows the result when the interaction (moiré confinement) strength is multiplied by a factor $x_i$ ($x_m$). For both $x_i \neq 1$ and $x_m \neq 1$, the reduction of the peak intensity is accompanied by the blue shift of its energy, consistent with the experimental observation in Fig. 3b and 3c. The much smaller energy shift of peak M than that of N under $x_i \neq 1$ is also consistent with the experiment. We therefore attribute the observed variations of peak intensities and energies in Fig. 3 to be mainly induced by the electric field modulation to the interaction strength, combined with some small contribution from the modulation to the moiré confinement.

4) The simulated polar plot of the total Raman intensity $I_\parallel + I_\perp$ indicates pronounced in-plane anisotropy under various linearly polarized excitations, see the inset in the right panel of Fig. 4d.

5) Our simulation well reproduces the magnetic field induced rotation of the anisotropy axis in the polar plot of $I_\parallel + I_\perp$, see Fig. 4e. Note that the magnetic field can introduce a valley splitting to the exciton[47] and change the dispersion/polarization of the electronic phonon modes. In our simulation, we find the rotation is majorly induced by the former.

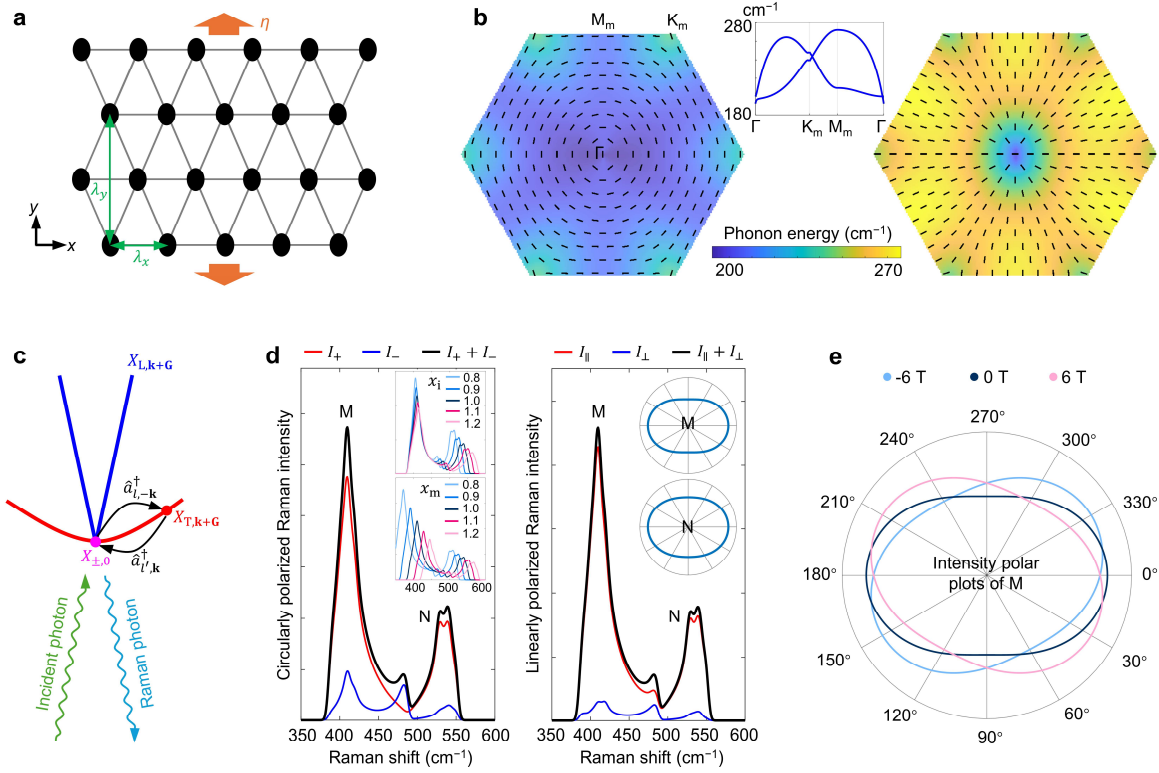

**Fig. 4 | Theoretical model for the electronic phonons in the electron crystal. a**, Schematic of a triangular electron crystal expanded along $y$ direction by a small fraction $\eta$, with $\lambda_y/\lambda_x = \sqrt{3}(1+\eta)$. **b**, The calculated dispersions and polarizations of the lower (left panel) and upper (right panel) electronic phonon branches across the moiré Brillouin zone. The middle inset shows the dispersion linecut along $\Gamma$-$K_m$-$M_m$-$\Gamma$. **c**, Schematic of the exciton-mediated two-phonon Raman process. Red and blue lines represent the transverse ($X_{T,\mathbf{k}+\mathbf{G}}$) and longitudinal ($X_{L,\mathbf{k}+\mathbf{G}}$) exciton branch, respectively, with $\mathbf{k}+\mathbf{G}$ the exciton center-of-mass momentum. Black arrows denote the two-step process $|X_{\pm,0}\rangle \to |X_{T,\mathbf{k}+\mathbf{G}}\rangle \otimes \hat{a}^\dagger_{l,-\mathbf{k}}|0\rangle \to |X_{\pm,0}\rangle \otimes \hat{a}^\dagger_{l',\mathbf{k}}\hat{a}^\dagger_{l,-\mathbf{k}}|0\rangle$. $\hat{a}^\dagger_{l,\mathbf{k}}$ is the creation operator of the $l$-th phonon branch with a wave vector $\mathbf{k}$. The bright excitons $|X_{\pm,0}\rangle$ in $\pm\mathbf{K}$ valleys mediate the incident photon absorption and Raman photon emission. **d**, Simulated polarization-resolved Raman spectra under circularly (left panel) and linearly (right panel) polarized excitations in the absence of a magnetic field. The linear polarization of the incident light is set along $x$. Red (blue) curves show the co-polarized (cross-polarized) Raman spectra, and black curves indicate the total intensities. The upper (lower) inset in the left panel shows the variation of $I_+ + I_-$ when the electron-electron interaction (moiré confinement) strength is multiplied by a factor $x_i$ ($x_m$). Insets in the right panel show polar plots of $I_\parallel + I_\perp$ as functions of the incident polarization direction, displaying in-plane anisotropy for both M and N. **e**, Simulated polar plots of $I_\parallel + I_\perp$ for peak M under magnetic fields ranging from −6 to 6 T.

**Conclusion**

Our spectroscopic observation of the electrically and magnetically tunable electronic phonons in a moiré-trapped electron crystal unveils a highly programmable quantum solid governed by strong electronic correlations. These electronic phonons, which emerge from long-range Coulomb interactions rather than atomic lattice vibrations, represent a new class of collective excitations in two-dimensional materials. The unraveled pronounced resonant enhancement under the excitonic excitation highlights the complex interplay between correlated insulators and free quasiparticles in moiré systems. More importantly, the high tunability of the electron crystal via external fields distinguishes these electronic phonons from atomic lattice phonons, providing a powerful spectroscopic window into the dynamics of moiré-induced correlated electronic orders.

Looking ahead, electronic phonons are also expected to occur in other fractionally filled correlated insulators observed in moiré superlattices. In particular, such modes in the honeycomb-type electron crystal at $v = 2/3$ are found to act differently from those in the triangular-type crystal discussed here[11]. The circular motions of electron sites and the resulting chiral phonons in the honeycomb crystal can bring exotic responses to external magnetic fields and interactions with excitons, which deserve further study. In the present experiments, such electronic phonons at this fractional filling are not resolved, likely reflecting the reduced stability and spectroscopic visibility of the corresponding electron crystal, which can be more sensitive to disorder, strain, and screening effects at certain fractional fillings. Future studies with optimized device parameters and enhanced experimental sensitivity will be important for a systematic investigation of electronic phonons at additional fractional fillings. The electronic phonons can also couple to various quantum degrees-of-freedom of the electron, e.g., spin through the spin-orbit interaction and layer-pseudospin through the position-dependent interlayer hopping, which may inspire novel methods for manipulating these degrees-of-freedom. More broadly, our work may motivate more theoretical and experimental research towards other kinds of collective quasiparticles in 2D moiré superlattices, such as the plasmons from charge excitations, magnons in ferromagnetic phases, and magnetorotons in fractional quantum Hall states.

**Methods**

**Sample fabrication**

For the dual-gate devices, thin hBN and graphite flakes were used as gate dielectric layers and gate electrodes, respectively. All the 2D material flakes were mechanically exfoliated from the bulk crystals and stacked layer-by-layer using a polypropylene carbon (PPC)-assisted dry-transfer method. The heterostructures were finally transferred onto a $SiO_2$/Si substrate with pre-patterned Au/Cr electrodes. The thicknesses of hBN flakes were characterized by atomic force microscopy (AFM) (see Supplementary Fig. 11, 12).

**Low-temperature Raman and PL spectroscopy**

The optical measurements were performed in a home-made confocal optical system integrated with a closed-cycle cryostat (Montana Instruments), enabling stable temperature control down to 5 K. A 632.8 nm He-Ne laser was used for Raman and PL measurements. For excitation-energy-dependent measurements, a tunable laser from continuous-wave optical parametric oscillator (C-Wave OPO, HÜBNER Photonics) provided variable-wavelength laser excitation. The laser power was kept in the range of 5~10 μW. The incident laser was focused on the sample by a 100× objective. The reflected light was collected in the backscattering configuration and detected by a spectrometer (Andor Shamrock 500i) for Raman spectra or an avalanche photodiode (APD) for IX PL intensity mapping. For IX PL measurements, an 850 nm long-pass filter was used. For Raman measurements, a 633 nm long-pass edge filter and a 1200 lines/mm grating were used. The magnetic field-related measurements were performed in attoDRY 2100 cryostat equipped with superconducting magnet.

**Polarization-resolved measurements**

We adopted two configurations to measure the polarization-resolved spectra. In the first configuration, aimed at analyzing the polarization state of Raman or PL emission, a fixed polarizer was inserted in the incident beam path to define the input polarization (linear or circular). For linear polarization, the scattered light was passed through a rotatable analyzer to acquire polarization-resolved intensities. For circular polarization, one quarter-wave plate was put after the polarizer in the incident light path to generate circularly polarized light, with the reflected light analyzed via the same quarter-wave plate followed by a linear analyzer. In the second configuration which was used to probe the in-plane anisotropy of the Raman modes, one polarizer was used to generate linearly polarized light, after which one half-wave plate was used to rotate the incident polarization direction continuously. The scattered light was collected without an analyzer, and the total intensity was recorded as a function of the incident polarization angle. The measured intensity polar plots reflect whether there exists anisotropy for Raman or PL signals.

**Linear dichroism (LD) measurement**

For LD measurement, a 632.8 nm linearly polarized laser was first modulated by a mechanical chopper and then passed through a photo-elastic modulator (PEM, Hinds Instruments). The fast axis of the PEM was oriented 45° with respect to the incident polarization. The PEM introduced a half-wave retardance ($\lambda/2$) at its modulation frequency, alternating the incident polarization between two orthogonal directions. The laser was focused on the sample at normal incidence and the reflected light was measured by a photodiode (PD). The signal was demodulated by two lock-in amplifiers (Zurich Instruments MFLI). One was referenced to the PEM modulation frequency to get the reflectance difference for incident light in two orthogonal polarization directions and the other one was referenced to the working frequency of chopper to get the total reflectance.

**Details of the theoretical model**

We simulate the Raman-active collective modes based on a theory of the interaction between the exciton and electronic phonon modes. The exciton Hamiltonian under the electron-hole exchange interaction and an out-of-plane magnetic field in monolayer transition metal dichalcogenides (TMDs) is:

$$\hat{H}_X = \sum_{\mathbf{Q}} (E_{X,\mathbf{Q}} + |J_\mathbf{Q}|)(|X_{+,\mathbf{Q}}\rangle\langle X_{+,\mathbf{Q}}| + |X_{-,\mathbf{Q}}\rangle\langle X_{-,\mathbf{Q}}|)$$

$$+ \frac{1}{2}\Delta E_B \sum_{\mathbf{Q}} (|X_{+,\mathbf{Q}}\rangle\langle X_{+,\mathbf{Q}}| - |X_{-,\mathbf{Q}}\rangle\langle X_{-,\mathbf{Q}}|) \quad (1)$$

$$+ \sum_{\mathbf{Q}} (J_\mathbf{Q}|X_{+,\mathbf{Q}}\rangle\langle X_{-,\mathbf{Q}}| + \text{h.c.}).$$

Here $|X_{\pm,\mathbf{Q}}\rangle$ is the intravalley exciton in $\pm\mathbf{K}$ valley with a center-of-mass momentum $\mathbf{Q} \equiv (Q_x, Q_y) = Q(\cos\theta_\mathbf{Q}, \sin\theta_\mathbf{Q})$, $\Delta E_B$ is the valley splitting induced by the magnetic field. Here we only consider the lowest-energy 1s exciton, the other Rydberg states (2p, 3d, …) have ~130 meV higher energies thus are dropped. The strength of the electron-hole exchange interaction has the form $J_\mathbf{Q} \approx J_{ex} Q e^{-2i\theta_\mathbf{Q}}$ with $J_{ex} \approx 1$ eV·Å[48]. The exciton eigenstates $|X_{1,\mathbf{Q}}\rangle$ and $|X_{2,\mathbf{Q}}\rangle$ are solved from Eq. (1), which under $\Delta E_B = 0$ become the transverse branch $|X_{T,\mathbf{Q}}\rangle \equiv \frac{1}{\sqrt{2}}(e^{-i\theta_\mathbf{Q}}|X_{+,\mathbf{Q}}\rangle - e^{i\theta_\mathbf{Q}}|X_{-,\mathbf{Q}}\rangle)$ and longitudinal branch $|X_{L,\mathbf{Q}}\rangle \equiv \frac{1}{\sqrt{2}}(e^{-i\theta_\mathbf{Q}}|X_{+,\mathbf{Q}}\rangle + e^{i\theta_\mathbf{Q}}|X_{-,\mathbf{Q}}\rangle)$ (see Fig. 4c). Note that the electrostatic potential generated by the anisotropic electron crystal can introduce spatial anisotropy to the exciton, which is simply modeled by different effective masses $M_x \neq M_y$ in the dispersion relation $E_{X,\mathbf{Q}} \equiv E_{X,0} + \frac{\hbar^2 Q_x^2}{2M_x} + \frac{\hbar^2 Q_y^2}{2M_y}$. In our simulation we set $M_x = 1.05 m_0$ and $M_y = 1.65 m_0$.

The electronic phonon mode is described by the Hamiltonian

$$\hat{H}_{\text{ph}} = \sum_{l=1}^{2} \sum_{\mathbf{k}} \hbar \omega_{l,\mathbf{k}} \left( \hat{a}_{l,\mathbf{k}}^{\dagger} \hat{a}_{l,\mathbf{k}} + \tfrac{1}{2} \right). \tag{2}$$

Here, $\hat{a}_{l,\mathbf{k}}$ is the annihilation operator of the $l$-th phonon branch with a wave vector $\mathbf{k}$ and frequency $\omega_{l,\mathbf{k}}$ ($l = 1, 2$). The interaction Hamiltonian between such an electronic phonon mode and the exciton is

$$\hat{H}_{\text{X-ph}} = \sum_{l,\mathbf{k}} (\hat{a}_{l,-\mathbf{k}}^{\dagger} + \hat{a}_{l,\mathbf{k}}) \sum_{\mathbf{Q},\mathbf{G}} \alpha_{l,\mathbf{k}+\mathbf{G}} (|X_{+,\mathbf{Q}+\mathbf{k}+\mathbf{G}}\rangle\langle X_{+,\mathbf{Q}}| + |X_{-,\mathbf{Q}+\mathbf{k}+\mathbf{G}}\rangle\langle X_{-,\mathbf{Q}}|). \tag{3}$$

Here $\alpha_{l,\mathbf{k}+\mathbf{G}} \propto \frac{\mathbf{A}_{l,\mathbf{k}}^{*} \cdot (\mathbf{k}+\mathbf{G}) V(\mathbf{k}+\mathbf{G})}{\sqrt{\omega_{l,\mathbf{k}}}} \left[ \left(1 + \left(\frac{m_h a_B |\mathbf{k}+\mathbf{G}|}{m_e + m_h}\right)^2\right)^{-3/2} - \left(1 + \left(\frac{m_e a_B |\mathbf{k}+\mathbf{G}|}{m_e + m_h}\right)^2\right)^{-3/2} \right]$ is the interaction strength (see Supplementary Section 4 for a detailed derivation), with $\mathbf{A}_{l,\mathbf{k}}$ the normalized phonon polarization vector, $\mathbf{G}$ the moiré reciprocal lattice vector, $V(\mathbf{k}+\mathbf{G})$ the $\mathbf{k}$-space Coulomb potential form, $m_{e/h}$ the electron/hole effective mass and $a_B \approx 2$ nm the exciton Bohr radius[49]. The interaction Hamiltonian $\hat{H}_{\text{X-ph}}$ then leads to phonon-assisted intravalley scatterings for the exciton.

We expect the photon absorption and emission in the Raman process involving electronic phonons to be mediated by the bright exciton $|X_{\pm,0}\rangle$. Given the inversion symmetry of the electron crystal, the phonon at $\mathbf{k} = 0$ has an odd parity and should be Raman-inactive. This can also be seen from the fact that the exciton-phonon interaction strength $\alpha_{l,\mathbf{k}}$ vanishes at $\mathbf{k} = 0$. Therefore, we attribute the collective modes M and N to exciton-mediated two-phonon emissions (see Fig. 4c). The Raman intensity can be calculated from the fourth-order Fermi golden rule

$$I(\omega) \propto \sum_{f} \left| \sum_{\psi_1 \psi_2 \psi_3} \frac{M_{i \to \psi_1} M_{\psi_1 \to \psi_2} M_{\psi_2 \to \psi_3} M_{\psi_3 \to f}}{(E_i - \tilde{E}_{\psi_1})(E_i - \tilde{E}_{\psi_2})(E_i - \tilde{E}_{\psi_3})} \right|^2 \delta(E_i - E_f). \tag{4}$$

Here $E_i$ is the energy of the incident photon, and $E_f$ is that of the final state which consists of a Raman photon and two electronic phonons with opposite wave vectors. $\omega = \omega_{l,-\mathbf{k}} + \omega_{l',\mathbf{k}}$ is the Raman shift with $\omega_{l,-\mathbf{k}}$ and $\omega_{l',\mathbf{k}}$ the emitted electronic phonon energies. $\tilde{E}_{\mu} \equiv E_{\mu} + i\gamma_{\mu}$ with $E_{\mu}$ and $\gamma_{\mu}$ the energy and linewidth, respectively, for the intermediate state $\mu = \psi_1, \psi_2, \psi_3$. $\psi_1 = |X_{\pm,0}\rangle$ is a bright exciton; $\psi_2 = |X_{1,\mathbf{k}+\mathbf{G}}\rangle \otimes \hat{a}_{l,-\mathbf{k}}^{\dagger}|0\rangle$ or $|X_{2,\mathbf{k}+\mathbf{G}}\rangle \otimes \hat{a}_{l,-\mathbf{k}}^{\dagger}|0\rangle$ consists of an exciton at $\mathbf{k} + \mathbf{G}$ and an $l$-th branch electronic phonon at $-\mathbf{k}$, with $|0\rangle$ the phonon vacuum; $\psi_3 = |X_{\pm,0}\rangle \otimes \hat{a}_{l',\mathbf{k}}^{\dagger} \hat{a}_{l,-\mathbf{k}}^{\dagger}|0\rangle$ consists of a bright exciton and two electronic phonons with opposite wave vectors. $M_{\mu \to \nu}$ are matrix elements for virtual processes between states $\mu$ and $\nu$ induced by $\hat{H}_{\text{X-ph}}$.

Eq. (4) is used to calculate the two-phonon Raman intensities $I(\omega)$ under various excitation polarizations, with the simulation results shown in Fig. 4d and 4e. In our simulation, we set the energy of the incident photon to be the same as the exciton energy, $\gamma_{\psi_1} = \gamma_{\psi_3} = 1$ meV due to the large radiative decay rate of $|X_{\pm,0}\rangle$, and $\gamma_{\psi_2} = 0.1$ meV since $\psi_2$ involves dark excitons with $\mathbf{k} + \mathbf{G} \neq 0$. The $\delta$-function in Eq. (4) is approximated by a gaussian form $\delta(E_i - E_f) \approx \frac{1}{\sqrt{\pi}\sigma} e^{-(E_i - E_f)^2/\sigma^2}$ with $\sigma = 0.5$ meV.

**Methods references**

**Data availability**

Other data that support the findings of this study are available from the corresponding author upon request. Source data are provided with this paper.

**Acknowledgements**

We acknowledge helpful discussions with Ataç Imamoğlu, Yoshinori Tokura, Mohammad Hafezi and Lingjie Du. This work is supported by ASTAR (M21K2c0116, M24M8b0004), Singapore National Research foundation (NRF-CRP22-2019-0004, NRF-CRP30-2023-0003, NRF-CRP31-0001, NRF2023-ITC004-001 and NRF-MSG-2023-0002) and Singapore Ministry of Education Tier 2 Grant (MOE-T2EP50221-0005, MOE-T2EP50222-0018). X.C. acknowledges the support from the NTU Presidential Postdoctoral Fellowship (Grant No. 03INS001828C230). K.W. and T.T. acknowledge support from the JSPS KAKENHI (Grant Numbers 21H05233 and 23H02052), the CREST (JPMJCR24A5), JST and World Premier International Research Center Initiative (WPI), MEXT, Japan. H.Y. acknowledges support by NSFC under grant No. 12274477.

**Author contributions**

W.G. and Y.Z. conceived the project. Y.Z. and X.C. performed most of the experimental measurements. X.C., H.C., S.Y. and H.P. fabricated the heterostructure device. S.R., X.D., Y.Z., A.R. and Q.S. contributed to the discussion on experimental design and assisted the optical measurements. H.Y. and Y.H. performed theoretical calculations. K.W. and T.T. grew the hBN single crystal. Y.Z., H.Y. and W.G. analyzed the data and wrote the manuscript. All authors participated in discussions about the results.

**Competing interests**

The authors declare no competing interests.

**Additional information**

**Supplementary information.** The online version contains supplementary material.

**Correspondence and requests for materials** should be addressed to Hongyi Yu and Weibo Gao.

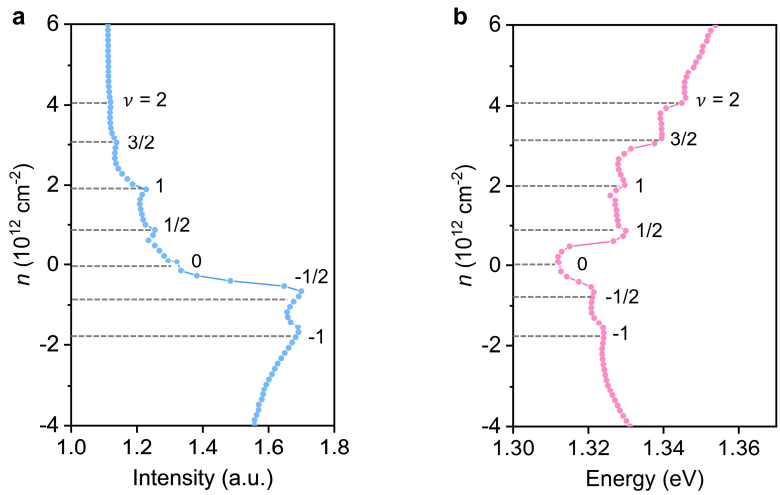

**Extended Data Fig. 1 | Assignment of filling factors using IX PL. a**, IX PL intensity as a function of carrier density, obtained from the data shown in Fig. 1b. Enhanced IX PL intensity is observed at specific carrier densities corresponding to correlated electronic states. **b**, Extracted IX PL peak energy as a function of carrier density. Clear blueshifts appear at the same filling factors as in (**a**).

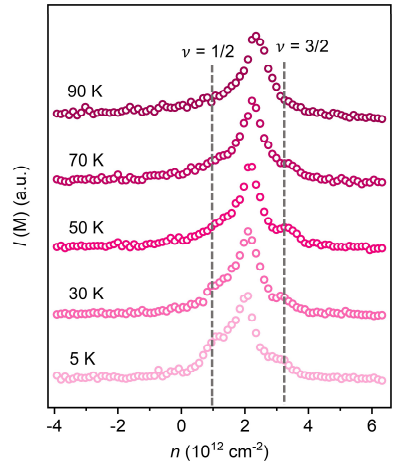

**Extended Data Fig. 2 | Temperature dependence of peak M at $v = 1/2$ and $v = 3/2$.** Intensity of peak M shows local maxima at fillings $v = 1/2$ and $v = 3/2$, indicating the emergence of this collective excitation in these correlated phases. The temperature-dependent measurements reveal that peak M at $v = 1/2$ and $v = 3/2$ gradually weakens and vanishes above ~50 K and ~90 K, respectively.

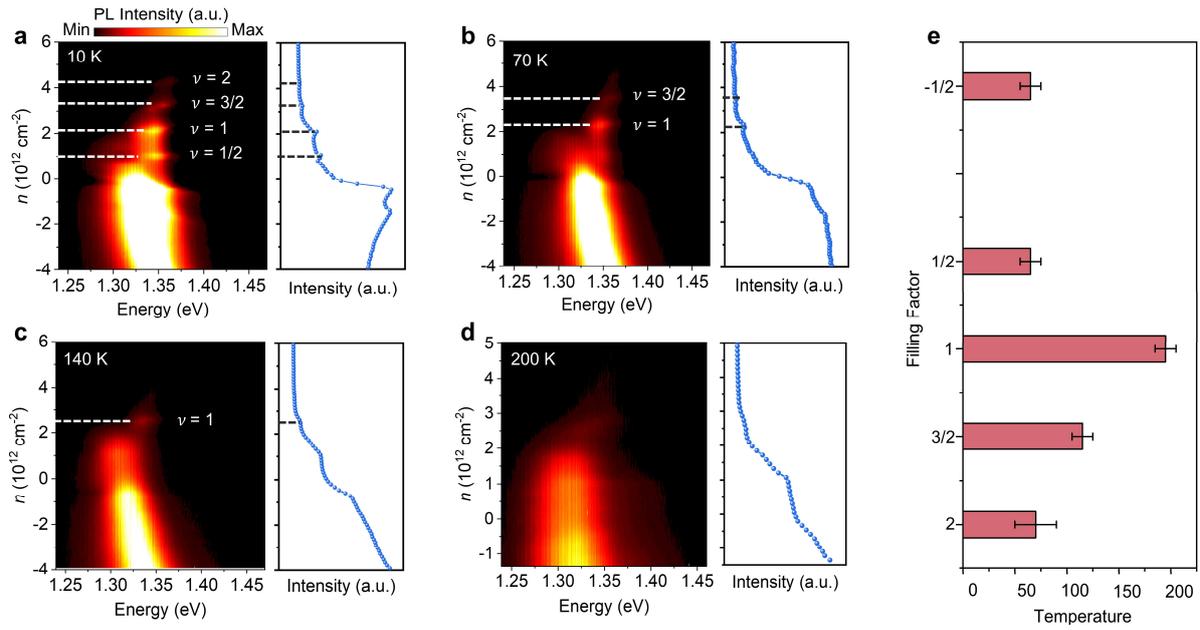

**Extended Data Fig. 3 | Temperature evolution of correlated phases revealed by IX PL. a-d**, PL spectra as a function of carrier density measured at 10 K (**a**), 70 K (**b**), 140 K (**c**) and 200 K (**d**). IX PL intensity is enhanced at specific fillings, as highlighted by the integrated intensity plot in right panel, indicating the presence of correlated insulating states. **e**, Critical temperatures of multiple correlated states assigned by the temperature dependence of IX PL. Error bars are the uncertainties of the extracted $T_c$ from repeated measurement cycles.

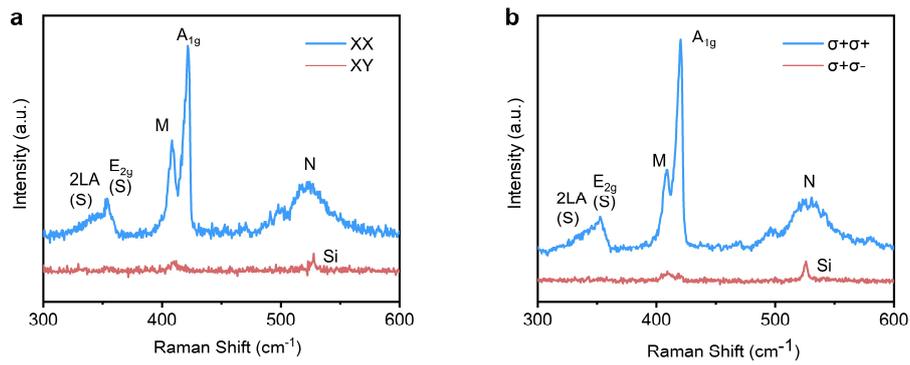

**Extended Data Fig. 4 | Polarization selection rules of peaks M and N. a**, Raman spectra measured under linearly polarized excitation in co-polarized (XX) and cross-polarized (XY) configurations. **b**, Raman spectra measured under circularly polarized excitation in co-polarized (σ+σ+) and cross-polarized (σ+σ-) configurations. The peak at ~520 cm$^{-1}$ is from silicon (Si) substrate, which is cross-polarized in both linear- and circular-polarization configurations, serving as a polarization reference.

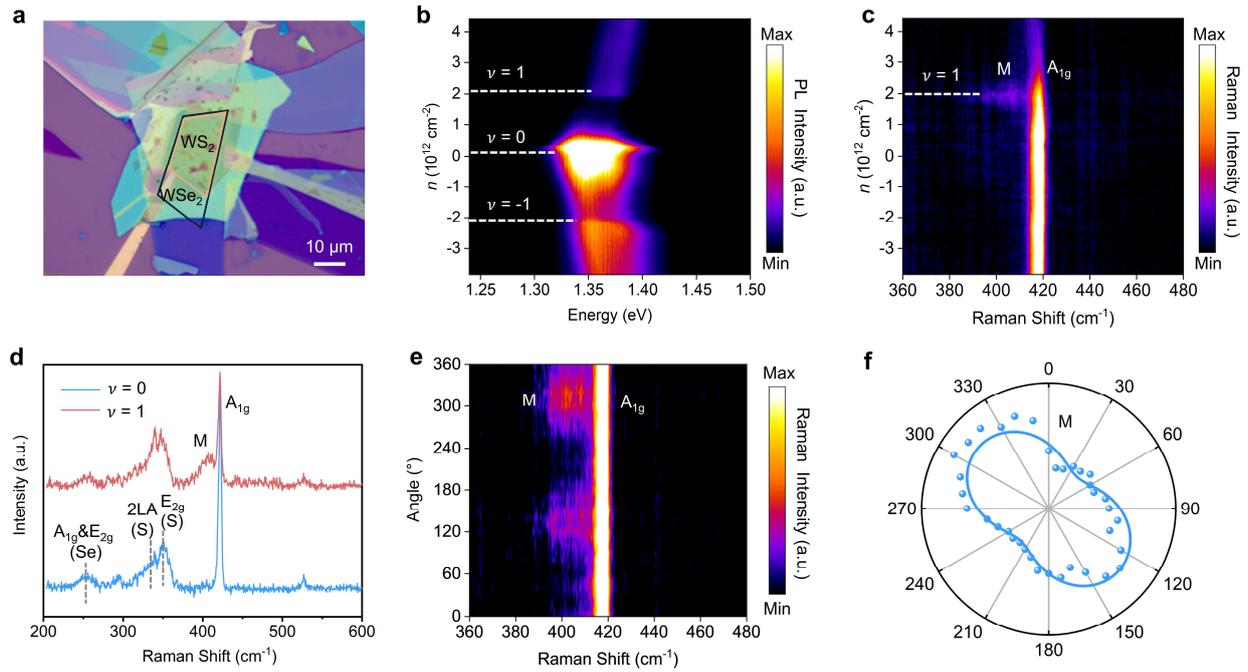

**Extended Data Fig. 5 | Observation of the new Raman mode in device D2. a**, Optical image of D2. The moiré density is ~2.03×10$^{12}$ cm$^{-2}$, corresponding to a twist angle of ~0.85°. **b**, Doping-dependent IX PL spectra, showing clearly resolved filling factors $\nu$ = –1, 0, and 1. **c**, Doping-dependent Raman spectra. Peak M emerges prominently near $\nu$ = 1. **d**, Extracted Raman spectra at $\nu$ = 0 and $\nu$ = 1 from (**c**). **e**, Raman spectra as a function of incident polarization angle (measured under configuration in Fig. 2f). **f**, Polar plot of the intensity of peak M, extracted from (**e**). The excitation laser energy is 1.96 eV.

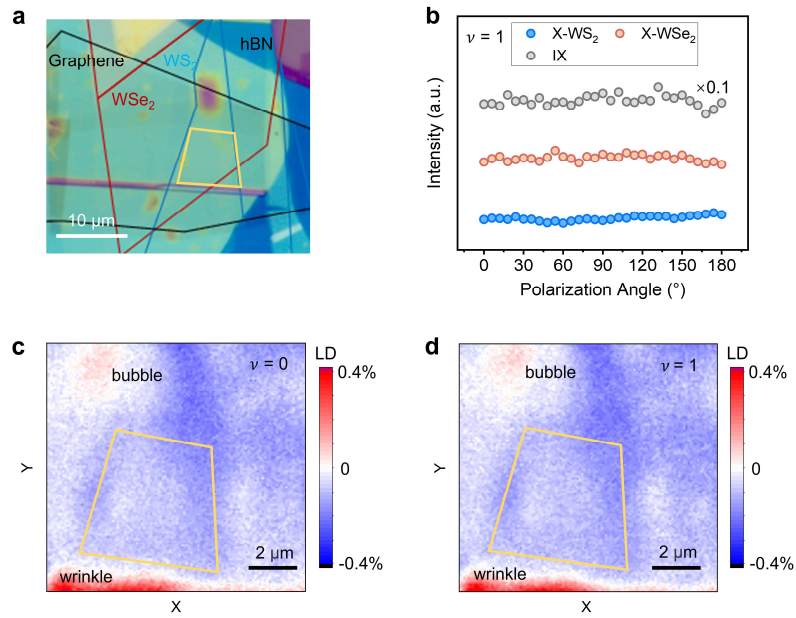

**Extended Data Fig. 6 | Polarization-dependent exciton PL and LD measurements. a**, Optical picture of the device. Yellow outlines mark the heterostructure region used for optical measurements in this study. **b**, Intralayer (X) and interlayer (IX) exciton PL intensity as a function of linear polarization angles of the incident light, revealing their isotropic responses. **c**, **d**, Spatial maps of LD signals measured at fillings $v = 0$ (**c**) and $v = 1$ (**d**). The heterostructure region shows negligible LD signal, in stark contrasts to that in the wrinkles and bubbles, indicating the high spatial uniformity.

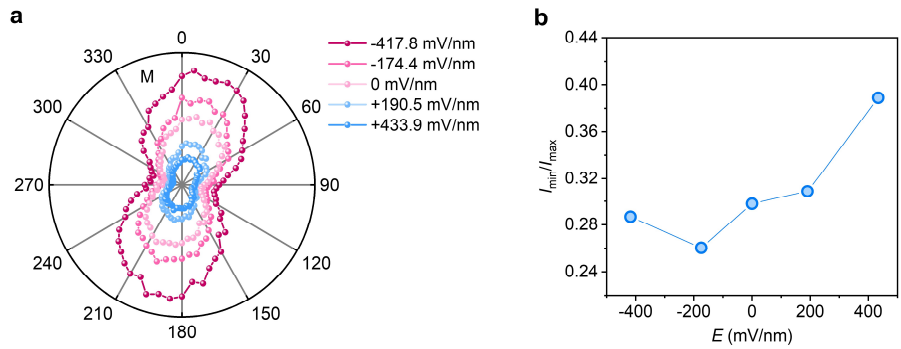

**Extended Data Fig. 7 | In-plane anisotropy of peak M under varying electric fields. a**, The intensity polar plots of M measured with varying incident linear polarizations at several electric field strengths. The orientation of anisotropy shows no obvious change across all field values. **b**, Extracted intensity ratios of the minimum to maximum values from the fitted polar plots in (**a**), representing the degree of in-plane anisotropy as a function of electric field. A higher ratio corresponds to reduced anisotropy, indicating that the anisotropy decreases with increasing electric field.

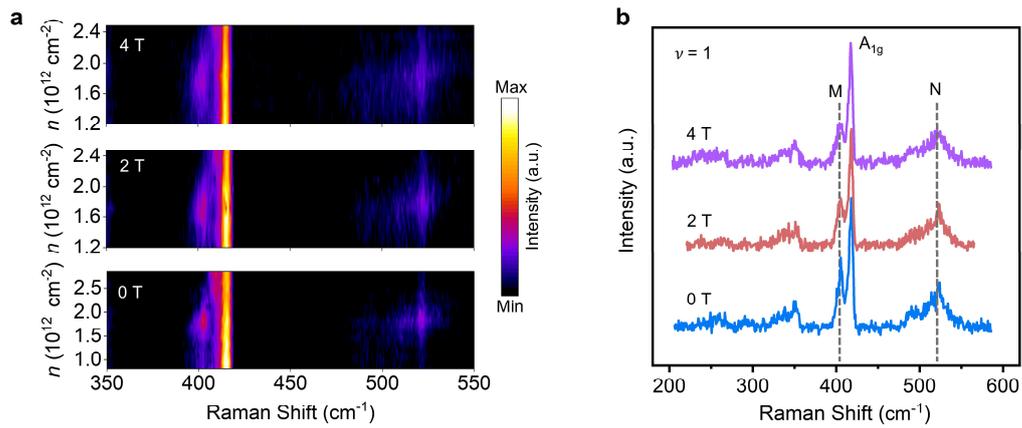

**Extended Data Fig. 8 | Magnetic field dependence of Raman spectra. a**, Carrier-density-dependent Raman spectra acquired under varying out-of-plane magnetic fields. **b**, Raman spectra at fixed filling $\nu = 1$ under variable magnetic fields. Peaks M and N exhibit negligible energy shifts.

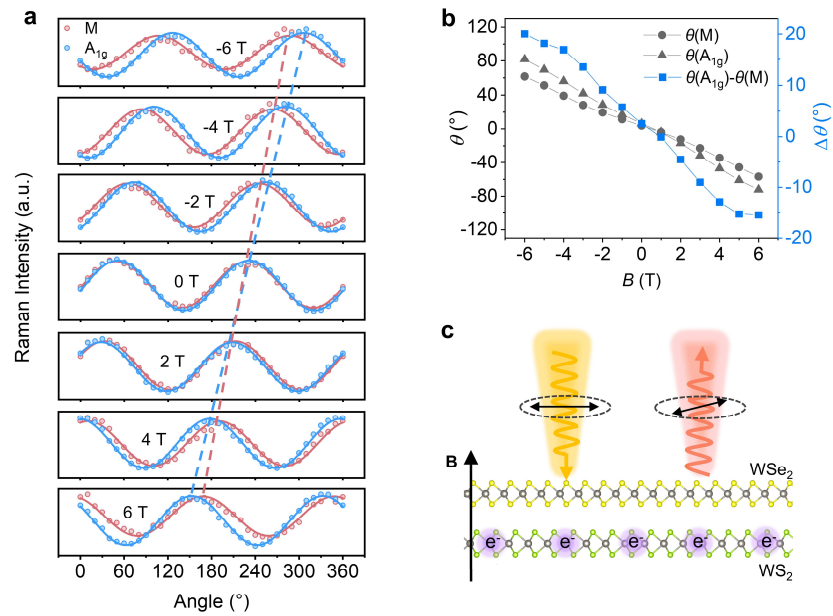

**Extended Data Fig. 9 | Magneto-optical effect of the electronic phonons. a**, Polarization states of M and $A_{1g}$ modes at $v = 1$ under varying magnetic fields. Dashed lines illustrate the angles of maximum Raman intensity, showing clear magnetic-field-induced polarization rotation. **b**, Fitted polarization phases ($\theta$) of M and $A_{1g}$ modes, along with their phase difference ($\Delta\theta$), as a function of magnetic field. The phase difference between M and $A_{1g}$ grows with increasing positive or negative magnetic field. Note that the exciton valley splitting induced by the applied magnetic field can lead to a rotation of the polarization of the scattered light. The distinct magnetic responses of M and $A_{1g}$ modes partly stem from the stronger resonant enhancement of the M mode by the exciton. The broken rotational symmetry of the electron crystal can also play a role here, which has been shown to affect the polarization angle (see Fig. 2g). **c**, Schematic of the polarization rotation of scattered light under an out-of-plane magnetic field.